\documentclass[12pt]{article}

\usepackage{sbc-template}
\usepackage{graphicx,url}
\usepackage[utf8]{inputenc}
\usepackage[brazil]{babel}
\usepackage[dvipsnames]{xcolor}
\usepackage{subcaption}
\usepackage{makecell}
\usepackage{longtable}
\usepackage{breakcites}
\usepackage{multirow}

\providecommand{\keywords}[1]{\textbf{\textit{Palavras-chave:}} #1}

\title{\textit{Guidelines} para Desenvolvimento de Jogos \textit{Mobile} Inclusivos}

\author{Gabriela Panta Zorzo, João Vitor Dall Agnol Fernandes, \\ Soraia Raupp Musse (Orientadora)}

\address{Escola Politécnica - Pontifícia Universidade Católica do Rio Grande do Sul \\
  Av. Ipiranga, 6681 – 90.619-900 – Porto Alegre – RS – Brasil}

\begin{document} 

\maketitle

\begin{resumo}
Os jogos representam uma parte significativa da cultura moderna, o que demonstra a importância de garantir que todas as pessoas possam participar e jogar para se sentirem incluídas na nossa sociedade. Porém, a maior parte dos jogos digitais acaba sendo inacessível para pessoas com alguma deficiência. Parte do problema na hora de pensar em um design de jogo inclusivo, é que não existe uma única solução para incluir acessibilidade, e o que funciona bem para um grupo pode não funcionar para outro. Este trabalho propõe um modelo de \textit{guidelines} para o desenvolvimento de jogos \textit{mobile} inclusivos, considerando o grande uso de \textit{smartphones} pela população e a necessidade de incluirmos pessoas com deficiência na cultura de jogos.
\end{resumo}


\section{Introdução}

Jogos digitais são extremamente comuns atualmente, e o número de pessoas tendo acesso a diferentes plataformas para jogar, especialmente \textit{smartphones}, tem crescido cada vez mais \cite{ULF}. Os jogos representam uma parte significativa da cultura moderna, o que demonstra a importância de garantir que todas as pessoas possam participar e jogar para se sentirem incluídas na nossa sociedade \cite{CAIRNS19}. A falta de jogos inclusivos faz com que as pessoas portadoras de alguma deficiência tenham dificuldades em serem inseridas dentro desse contexto social, o que pode fazer com que esse grupo acabe sendo excluído por não ter as mesmas oportunidades de participar deste meio e compartilhar essas experiências com as demais pessoas \cite{ULF}. 

Em muitos casos, a falta de um design inclusivo não acontece por uma escolha consciente, mas sim pela falta deste olhar na hora do desenvolvimento. Algumas adaptações no processo de \textit{game design} já podem ser o suficiente para tornar os jogos acessíveis para pessoas com deficiência, mas muitas vezes essas adaptações acabam não sendo feitas por não pensarem nesse público na hora do desenvolvimento. O pensamento de que "existem pessoas desenvolvendo jogos específicos para esse público" precisa ser evitado neste cenário, pois acaba gerando uma maior separação ao invés de inclusão \cite{MICHAELHERON}.

Para apoiar o desenvolvimento de jogos inclusivos, existem \textit{guidelines} propostas por diferentes autores como \cite{COGNITIVE} e \cite{ABLEGAMERS} com o intuito de orientar esse processo. Essas \textit{guidelines}, ou diretrizes, servem como um guia desde o momento de concepção do design do jogo até a etapa de testes após o desenvolvimento. Por se tratarem de orientações, é importante analisá-las levando em consideração o contexto específico em que são aplicadas, cabendo à equipe responsável determinar a melhor abordagem para atendê-las quando relevante.

Este trabalho traz estudos realizados nas áreas de jogos digitais inclusivos e diretrizes já existentes criadas por diferentes autores (Seção \ref{relacionados}). Com base nesses estudos, é feita uma proposta de um novo grupo de \textit{guidelines} adaptadas para o contexto de jogos para dispositivos móveis, categorizadas de acordo com o tipo de deficiência (Seção \ref{proposta}). Por fim, utiliza-se esse novo conjunto de diretrizes para avaliar cinco jogos selecionados, fornecendo os resultados dessa avaliação (ver Seção \ref{resultados}).

\section{Trabalhos Relacionados} \label{relacionados}

Assim como filmes, músicas e televisão, jogos são uma parte importante da cultura moderna. Diferente de outras formas de entretenimento, os jogos são ativos, isto é, nós precisamos interagir com eles para termos a experiência completa. Essa interação muitas vezes acontece através de uma combinação de ações com as mãos e reações rápidas a elementos visuais. Porém, a maior parte dos jogos digitais acaba sendo inacessível para pessoas com alguma deficiência. Isso acontece não pela dificuldade do jogo em si, mas pela falta de um design inclusivo que considere as necessidades dessas pessoas na hora de desenvolver um jogo \cite{MICHAELHERON}. 

O trabalho de \cite{CAIRNS21} fala sobre três fatores que motivam as pessoas a jogarem, sendo eles: a necessidade de autonomia, a necessidade de se sentir competente e a necessidade de estar conectado com outras pessoas. Ademais, além de atuar nessas necessidades psicológicas, os jogos podem melhorar o bem-estar dos jogadores. Socializar através de jogos faz parte de um contexto sociocultural muito importante, podendo ser uma ferramenta para inclusão. Contudo, jogos digitais dependem fortemente de componentes gráficos como principal forma de comunicação com os jogadores, seja para definir os personagens, criar o ambiente do jogo e mostrar as interações com a cena. Até mesmo o termo \textit{video game} mostra que esses jogos fazem parte de uma cultura visual dominante. Poucos jogos no mercado são acessíveis para jogadores com deficiência, e os que são, geralmente são desenvolvidos principalmente para aqueles com alguma deficiência visual, fazendo com que sejam menos atrativos para os demais jogadores e gerando uma maior separação ao invés de uma inclusão \cite{ULF}. 

Parte do problema na hora de pensar em um design de jogo inclusivo, é que não existe uma única solução para incluir acessibilidade, e o que funciona bem para um grupo pode não funcionar para outro. O trabalho de \cite{MICHAELHERON} traz princípios de design simples para serem considerados, como uso de \textit{voice over} em menus, \textit{feedbacks} sonoros e textuais para ambientar o jogador e informar ações, ajustes de tamanho de fonte e cuidado com uso de cores. Ele também sugere que algumas dessas configurações possam ser ativadas e desativadas de acordo com as preferências do usuário, tornando o jogo não apenas mais acessível para pessoas com deficiência, mas inclusivo para todas pessoas que optem por utilizar essas opções.

Outras práticas a serem aplicadas podem ser observadas nas \textit{guidelines} propostas no \textit{Game Accessibility Guidelines} \cite{GAG}, as quais estão divididas em três níveis que levam em consideração parâmetros como a quantidade de pessoas impactadas, a importância deste impacto na experiencia do jogo e o custo de implementação, sendo eles o nível básico, médio e avançado. As diretrizes ainda se encontram em categorias de tipo de deficiência abordada, sendo elas motora, cognitiva, visual, auditiva e de fala. Algumas convenções apresentadas, na sua maioria classificadas como básicas, são notadas em outros trabalhos já citados. Dentro dessa categoria, pode-se destacar exemplos como uso de linguagem simples e clara, garantir que nenhuma informação importante será transmitida apenas por cores ou por sons e garantir que as configurações são salvas. Outras \textit{guidelines} dessa categoria podem ser observadas na Tabela \ref{tab:guidelines_gag}.

\begin{small}
\begin{longtable}{|l|l|}
\caption{\textit{Guidelines} no nível básico propostas por \cite{GAG}}
\label{tab:guidelines_gag}\\
\hline
\textbf{\textit{Guideline}}                                                                                                                                    & \textbf{Deficiência} \\ \hline
\endhead
\makecell[l]{Permite que os controles sejam remapeados / reconfigurados.}                                                                                    & Motora                         \\ \hline
\makecell[l]{Certifica-se de que todas as áreas da interface do usuário possam ser acessadas \\ usando o mesmo método de entrada do jogo.}                    & Motora                         \\ \hline
\makecell[l]{Inclui uma opção para ajustar a sensibilidade dos controles.}                                                                                  & Motora                         \\ \hline
\makecell[l]{Garante que os controles sejam tão simples quanto possível ou fornece uma \\ alternativa mais simples.}                                          & Motora                         \\ \hline
\makecell[l]{Garante que os elementos interativos / controles virtuais sejam grandes e bem \\ espaçados, especialmente em telas pequenas ou sensíveis ao toque.} & Motora                         \\ \hline
\makecell[l]{Inclui \textit{toggle} para qualquer sensação tátil.}                                                                          & Motora                         \\ \hline
\makecell[l]{Permite que o jogo seja iniciado sem a necessidade de navegar por vários níveis \\ de menus.}                                                    & Cognitiva                      \\ \hline
\makecell[l]{Usa um tamanho de fonte padrão facilmente legível.}                                                                                          & Cognitiva                      \\ \hline
\makecell[l]{Usa uma linguagem simples e clara.}                                                                                                            & Cognitiva                      \\ \hline
\makecell[l]{Usa formatação simples de texto.}                                                                                            & Cognitiva                      \\ \hline
\makecell[l]{Inclui tutoriais interativos.}                                                                                                               & Cognitiva                      \\ \hline
\makecell[l]{Permite que os jogadores progridam através de avisos de \\ texto em seu próprio ritmo.}                                                          & Cognitiva                      \\ \hline
\makecell[l]{Evita imagens piscantes e padrões repetitivos.}                                                                                            & Cognitiva                      \\ \hline
\makecell[l]{Certifica-se de que nenhuma informação essencial seja transmitida apenas \\ por cor.}                                                        & Visual                         \\ \hline
\makecell[l]{Se o jogo usar campo de visão (somente mecanismo 3D), define um padrão \\ apropriado para o ambiente de visualização esperado.}                   & Visual                         \\ \hline
\makecell[l]{Evita gatilhos de enjoo em simulação de VR.}                                                                                                  & Visual                         \\ \hline
\makecell[l]{Usa um tamanho de fonte padrão facilmente legível.}                                                                                            & Visual                         \\ \hline
\makecell[l]{Usa formatação simples de texto.}                                                                                            & Visual                         \\ \hline
\makecell[l]{Fornece alto contraste entre texto / UI e plano de fundo.}                                                                                       & Visual                         \\ \hline
\makecell[l]{Garante que os elementos interativos / controles virtuais sejam grandes e bem \\ espaçados, especialmente em telas pequenas ou sensíveis ao toque.} & Visual                         \\ \hline
\makecell[l]{Fornece legendas para todos os discursos importantes.}                                                                                         & Auditiva                       \\ \hline
\makecell[l]{Fornece controles de volume separados ou silenciamentos para efeitos, fala e \\ fundo / música.}                                                    & Auditiva                       \\ \hline
\makecell[l]{Certifica-se de que nenhuma informação essencial seja transmitida apenas \\ por sons.}                                                           & Auditiva                       \\ \hline
\makecell[l]{Se forem usadas legendas, apresente-as de forma clara e fácil de ler.}                                                                & Auditiva                       \\ \hline
\makecell[l]{Certifica-se de que a entrada de voz não seja necessária e incluída apenas como \\ um método de entrada complementar / alternativo.}               & Fala                           \\ \hline
\makecell[l]{Oferece uma ampla escolha de níveis de dificuldade.}                                                                                           & Geral                          \\ \hline
\makecell[l]{Fornece detalhes dos recursos de acessibilidade na embalagem e / ou site.}                                                                       & Geral                          \\ \hline
\makecell[l]{Fornece detalhes dos recursos de acessibilidade no jogo.}                                                                                      & Geral                          \\ \hline
\makecell[l]{Certifica-se de que todas as configurações sejam salvas / lembradas.}                                                                           & Geral                          \\ \hline
\makecell[l]{Solicita \textit{feedback} sobre acessibilidade.}                                                                                                       & Geral                          \\ \hline
\end{longtable}
\end{small}

Organizações, como a \textit{AbleGamers Foundation}\footnote{https://ablegamers.org}, trabalham para que os jogos sejam desenvolvidos de maneira inclusiva para atender pessoas com diferentes tipos de deficiência, diminuindo o isolamento social dessas pessoas e melhorando a qualidade de vida. O guia \textit{Includification} \cite{ABLEGAMERS} foi pensado para auxiliar no processo de criação de jogos acessíveis. Nele são encontradas \textit{guidelines} que também consideram diferentes tipos de deficiência: motora, visual, auditiva e cognitiva. Ademais, ele traz uma sessão com foco específico para acessibilidade em jogos de dispositivos móveis, que estão descritas na Tabela \ref{tab:guidelines_includification} de maneira resumida. Essa sessão aborda diretrizes com foco no toque, botões alternativos, alto contraste, opções modo daltonismo e configuração de velocidade. Também é reforçado que além dessas diretrizes específicas para esse tipo de dispositivo, ainda precisam ser cuidadas as demais diretrizes quando forem relevantes.

\begin{table}[ht]
\caption{\textit{Guidelines mobile} do \textit{Includification} \cite{ABLEGAMERS}}
\label{tab:guidelines_includification}
\begin{tabular}{|l|l|}
\hline
\textbf{Categoria}                       & \textit{\textbf{Guideline}}                                                                                                                                               \\ \hline
Toque                                    & \makecell[l]{Grande área que pode ser considerada a área para \\ tocar na tela.}                                                                                                           \\ \hline
Multi-toque                              & \makecell[l]{Se o seu jogo exigir pressionar vários lugares ao \\ mesmo tempo, considere agrupá-los.}                                                                                      \\ \hline
Botões alternativos                      & \makecell[l]{Se um jogo exigir um recurso especial, como tocar na \\ parte traseira do dispositivo ou girá-lo, permita que \\ meios alternativos sejam usados para atingir o \\ mesmo objetivo.} \\ \hline
Alto contraste                           & \makecell[l]{Use cores que sejam fáceis de distinguir de outros \\ elementos ambientais do jogo.}                                                                                          \\ \hline
                                         & \makecell[l]{Opções de modo daltonismo que podem ser \\ habilitadas.}                                                                                                                      \\ \cline{2-2} 
\multirow{-2}{*}{Opções modo daltonismo}  & \makecell[l]{Incluir símbolos para distinguir elementos \\ visualmente.}                                                                                           \\ \hline
Configuração de velocidade              & Capacidade de desacelerar o jogo.                                                                                                                                         \\ \hline
\end{tabular}%
\end{table}

O trabalho de \cite{TRISNADOLI} propõe um modelo para avaliação de qualidade para jogos \textit{mobile}. A necessidade de aplicar um modelo específico para esse tipo de jogos é devido a diferenças significativas do \textit{hardware} onde eles estão sendo executados. Jogos de dispositivos móveis são diferentes de jogos de computador ou \textit{video games} por diversos motivos, incluindo questões de usabilidade, jogabilidade e mobilidade. As diretrizes propostas no modelo de \cite{TRISNADOLI} levam em conta fatores de usabilidade, flexibilidade de uso e segurança. Dentro de flexibilidade de uso, um dos pontos levantados é a acessibilidade com relação ao jogo poder ser utilizado de maneira flexível, isto é, possuir formas alternativas de jogar, se acomodar bem ao ambiente e ter opções de controle apropriadas e flexíveis.

Pensando no contexto de jogos acessíveis para dispositivos móveis, \cite{COGNITIVE} analisa diretrizes propostas no \textit{Includification} \cite{ABLEGAMERS}, no \textit{Game Accessibility Guidelines} \cite{GAG} e por outros autores para verificar quais delas podem ser aplicadas para jogos \textit{mobile} acessíveis para pessoas com deficiência cognitiva. Elas são divididas em três níveis (Tabela \ref{tab:guidelines_cognitive}) de acordo com a complexidade de aplicações e do benefício que elas trazem para as pessoas. Exemplos que se destacam são: lembrança de objetivos e controles durante o jogo, possibilidade de repetição e menus acessíveis. 

\begin{small}
\begin{table}[ht]
\centering
\caption{\textit{Guidelines} propostas por \cite{COGNITIVE}}
\label{tab:guidelines_cognitive}
\begin{tabular}{|l|l|}
\hline
\textit{\textbf{Guideline}}                    & \textbf{Nível}                          \\ \hline
Utiliza uma linguagem simples.                 & \multirow{7}{*}{Baixo (Bom)}      \\ \cline{1-1}
Fontes personalizáveis (cor, tamanhos).        &                                         \\ \cline{1-1}
Ligar / desligar elementos gráficos.           &                                         \\ \cline{1-1}
Legendas.                                      &                                         \\ \cline{1-1}
Progressão simples a difícil.                  &                                         \\ \cline{1-1}
Menus acessíveis.                              &                                         \\ \cline{1-1}
Modos \textit{sandbox}.                                 &                                         \\ \hline
Níveis de treinamento.                         & \multirow{8}{*}{Médio (Melhor)}   \\ \cline{1-1}
Lembrança dos objetivos durante o jogo.          &                                         \\ \cline{1-1}
Lembrança dos controles durante o jogo.          &                                         \\ \cline{1-1}
Configuração alternativa de arquivos de som.   &                                         \\ \cline{1-1}
Use recompensas visuais explícitas.            &                                         \\ \cline{1-1}
Possibilidade de repetição.                    &                                         \\ \cline{1-1}
Pausa enquanto o texto está sendo lido.        &                                         \\ \cline{1-1}
Salvar configurações.                          &                                         \\ \hline
Mira automática, capacidade de travar um alvo. & \multirow{5}{*}{Alto (Excelente)} \\ \cline{1-1}
Configurações de velocidade.                   &                                         \\ \cline{1-1}
Capacidade de \textit{voice over}.             &                                         \\ \cline{1-1}
Sensibilidade ajustável / tolerância a erros.  &                                         \\ \cline{1-1}
Passe automático.                              &                                         \\ \hline
\end{tabular}%
\end{table}
\end{small}

Enquanto o trabalho de \cite{COGNITIVE} utiliza uma equação para avaliar o quanto um jogo atende as diretrizes, somando 1 para cada diretriz atendida e gerando uma nota ao final, outros trabalhos, como o \textit{Includification} \cite{ABLEGAMERS} e o \textit{Game Accessibility Guidelines} \cite{GAG}, optam por uma análise mais qualitativa. O \cite{GAG} disponibiliza uma tabela\footnote{https://gameaccessibilityguidelines.com/excel-checklist-download/} para ser utilizada de \textit{checklist} ao fazer a análise. Essa tabela considera se a \textit{guideline} é relevante para a mecânica do jogo em questão, se foi implementada e notas adicionais que possam ser necessárias. 

Embora existam diversas diretrizes para auxiliar no desenvolvimento de jogos inclusivos, as diretrizes de \cite{GAG} abrangem um escopo amplo, sem considerar o tipo de plataforma onde o jogo será executado. Já as diretrizes de \cite{COGNITIVE} são propostas para o escopo de jogos para dispositivos móveis, mas pensando apenas em deficiência cognitiva. O \textit{Includification} de \cite{ABLEGAMERS} traz as \textit{guidelines mobile} em uma sessão do seu guia, porém nessa sessão ele aborda somente diretrizes específicas para dispositivos móveis, sendo necessário buscar diretrizes gerais em outras sessões e analisar quais são relevantes. Desta forma encontra-se a oportunidade de definição de um conjunto de \textit{guidelines} para o desenvolvimento de jogos \textit{mobile} inclusivos, onde são abrangidos diferentes tipos de deficiência.

\section{Metodologia} \label{proposta}

\subsection{Avaliação por \textit{guidelines} já existentes} \label{analise}

\begin{figure} []
   \begin{minipage}{0.5\textwidth}
     \centering
     \includegraphics[width=.5\linewidth]{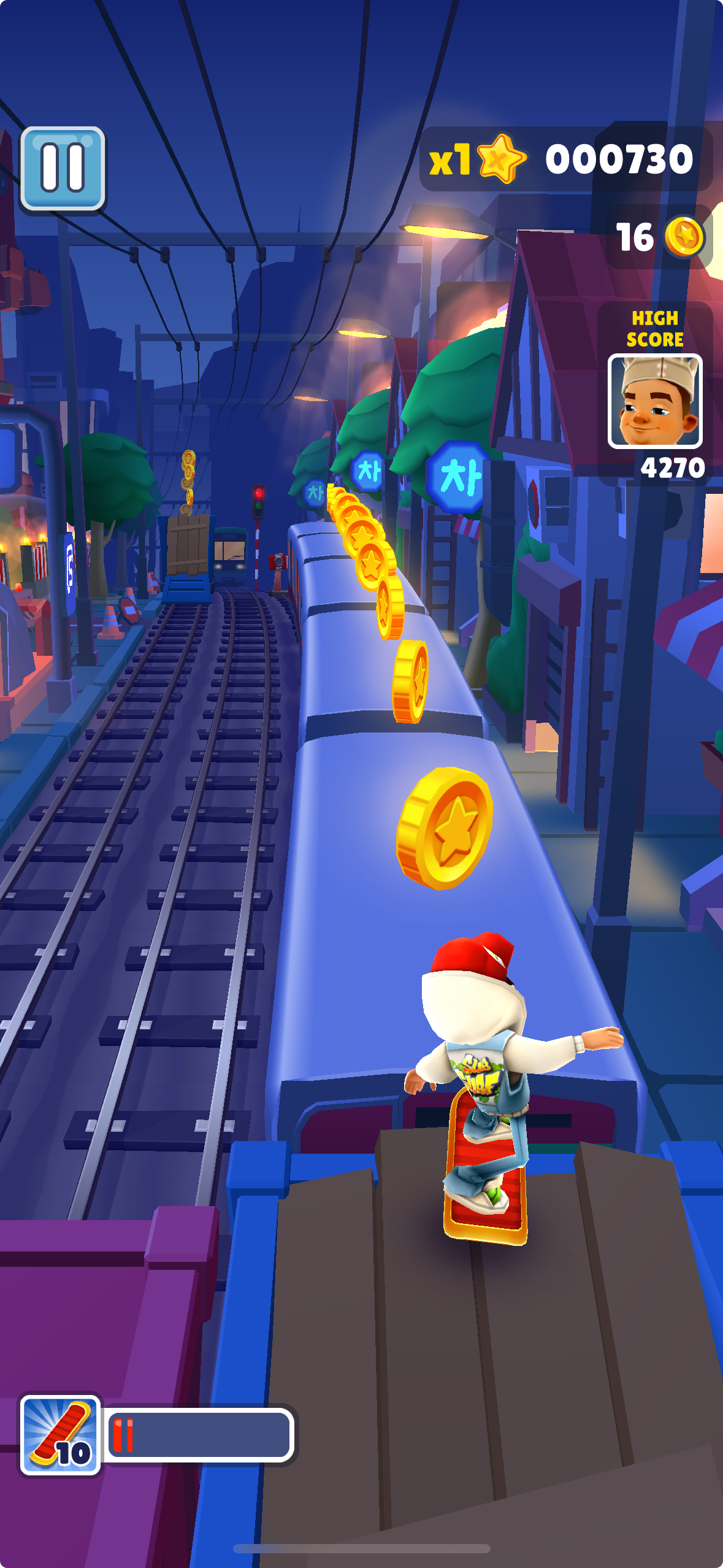}
     \caption{Subway Surfers}\label{fig:subway}
   \end{minipage}\hfill
   \begin{minipage}{0.5\textwidth}
     \centering
     \includegraphics[width=.5\linewidth]{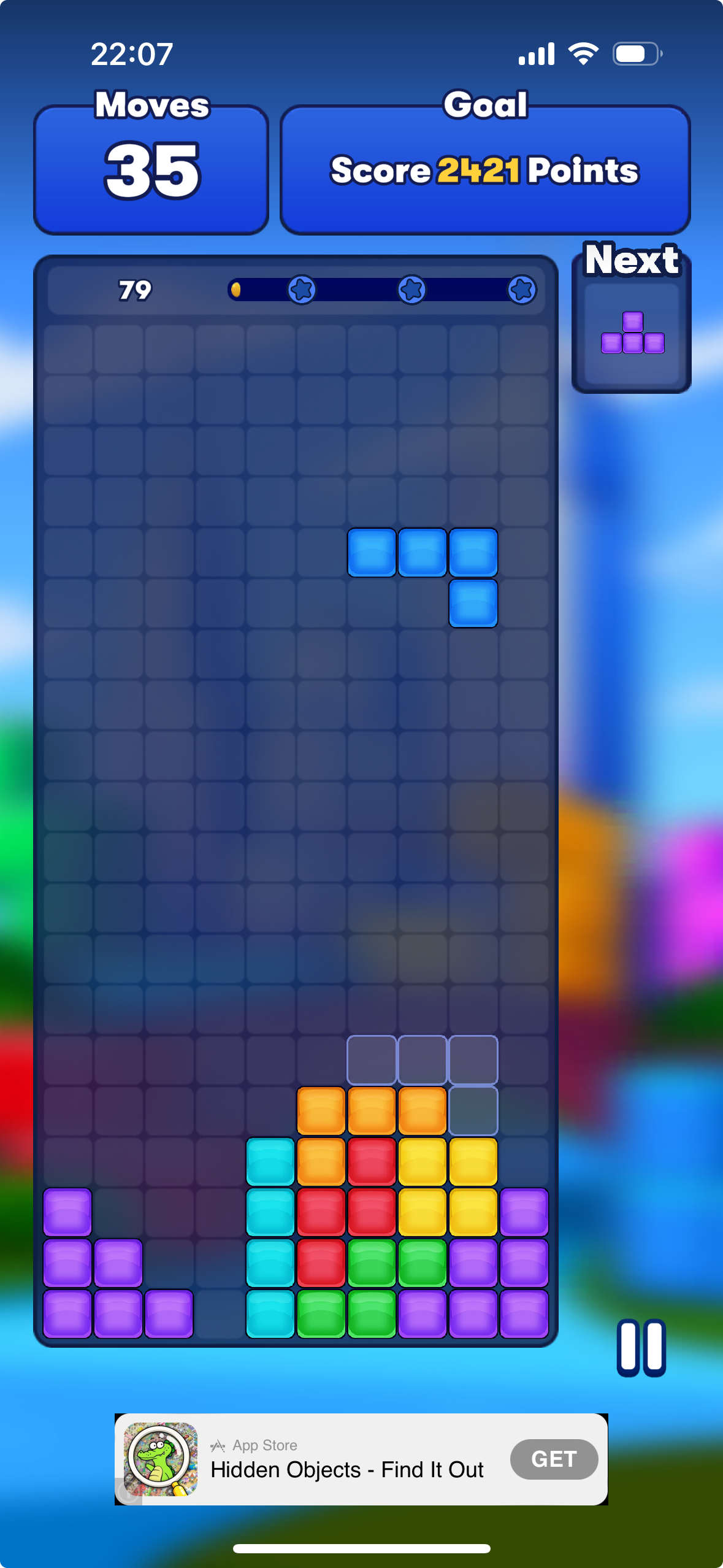}
     \caption{Tetris}\label{fig:tetris}
   \end{minipage}
\end{figure}

\begin{figure} 
     \centering
     \includegraphics[width= 0.59\linewidth]{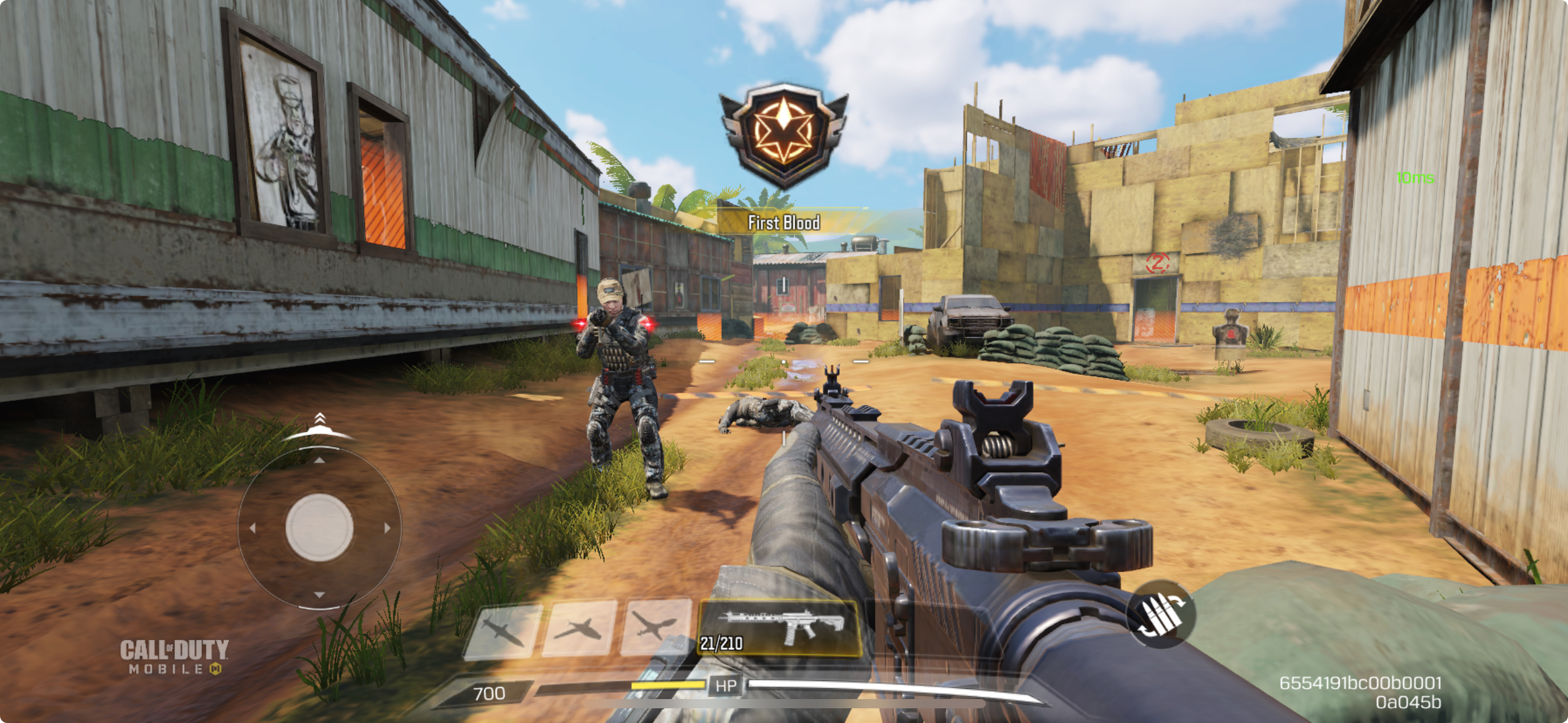}
     \caption{Call of Duty: Mobile}\label{fig:callofduty}
\end{figure}

\begin{figure} 
     \centering
     \includegraphics[width= 0.59\linewidth]{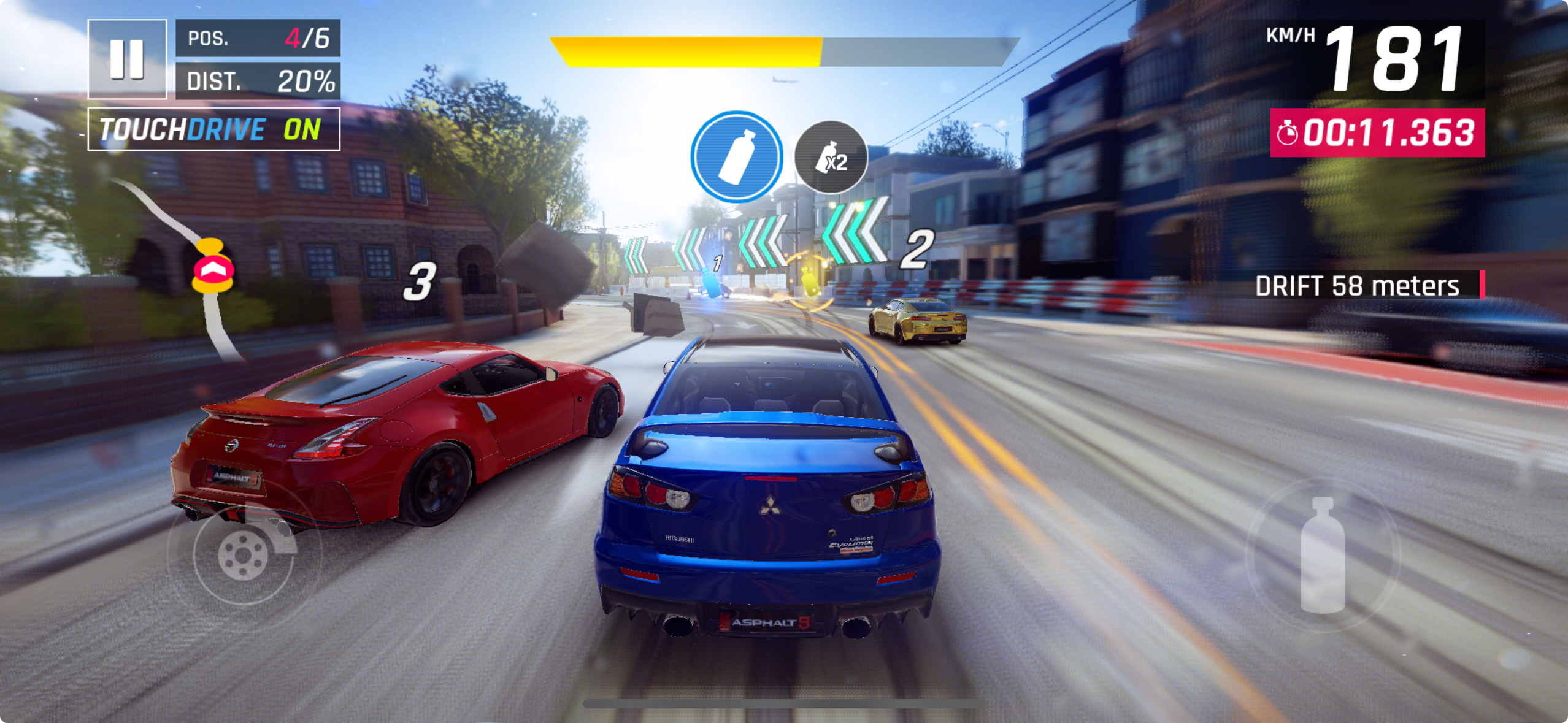}
     \caption{Asphalt 9}\label{fig:asphalt}
\end{figure}

\begin{figure}
    \centering
    \includegraphics[width= 0.59\linewidth]{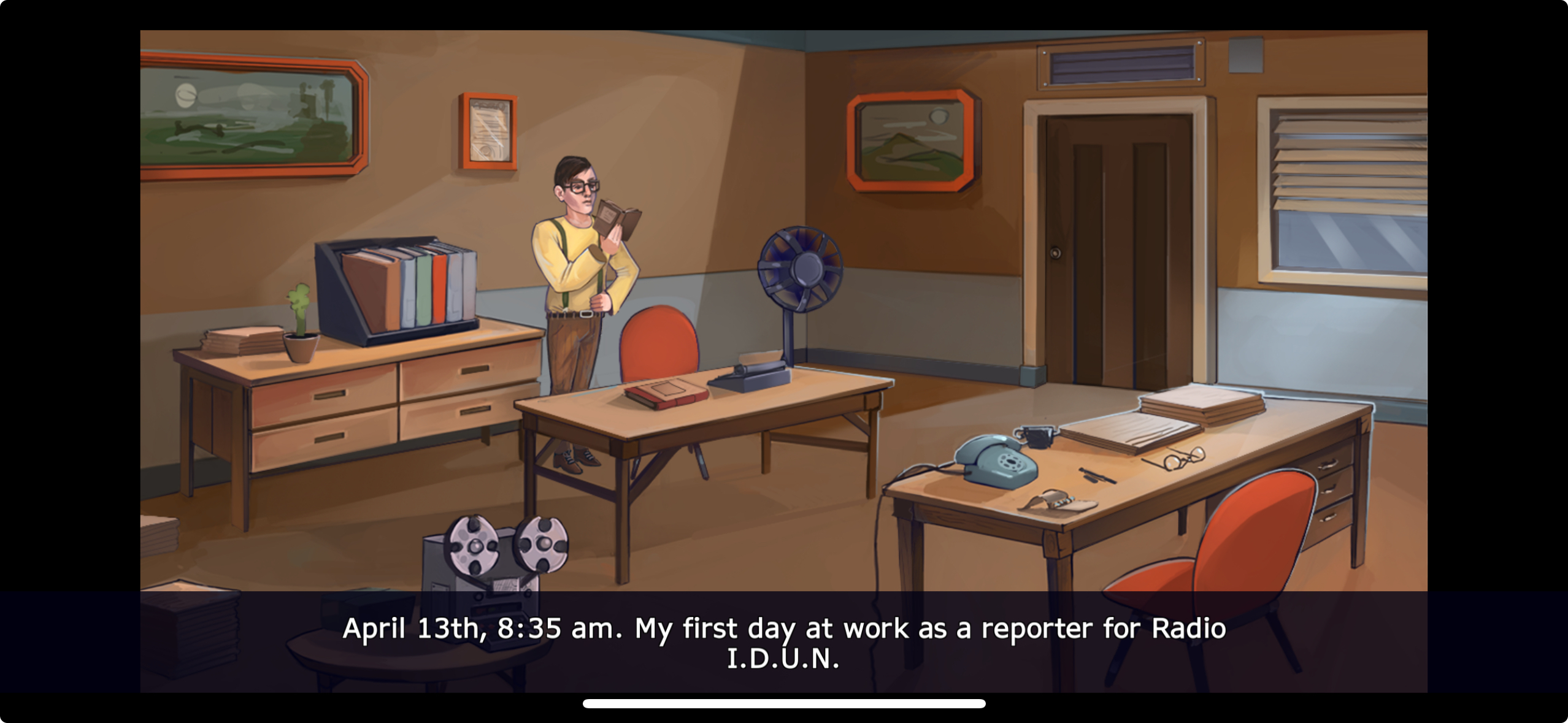}
    \caption{Frequency Missing}\label{fig:frequency}
\end{figure}

Com base nos trabalhos relacionados e tendo em vista o crescimento do número de dispositivos móveis, este trabalho propõe um modelo de \textit{guidelines} para o desenvolvimento de jogos \textit{mobile} inclusivos. Foram analisadas mais profundamente as diretrizes trazidas nos trabalhos de \cite{GAG}, \cite{ABLEGAMERS} e \cite{COGNITIVE} e vistas nas Tabelas \ref{tab:guidelines_gag}, \ref{tab:guidelines_includification} e \ref{tab:guidelines_cognitive} respectivamente. Para isso, foram escolhidos cinco jogos de categorias diferentes para serem avaliados de acordo com cada uma das diretrizes. Os jogos Subway Surfers\footnote{https://apps.apple.com/br/app/subway-surfers/id512939461} (Figura \ref{fig:subway}), Tetris\footnote{https://apps.apple.com/br/app/tetris/id1491074310} (Figura \ref{fig:tetris}), Call of Duty: Mobile\footnote{https://apps.apple.com/br/app/call-of-duty-mobile/id1287282214} (Figura \ref{fig:callofduty}) e Asphalt 9\footnote{https://apps.apple.com/br/app/asphalt-9-legends/id805603214} (Figura \ref{fig:asphalt}) foram escolhidos devido as suas classificações na App Store\footnote{https://www.apple.com/br/app-store/} no topo de cada uma das suas categorias. Além disso, cada um deles representa um tipo diferente de jogo, trazendo variedade para os tipos de comando utilizados e objetivos do jogo. São eles: \textit{runner}, retro, ação e corrida. Já o jogo Frequency Missing\footnote{https://apps.apple.com/br/app/frequency-missing/id1398341698} (Figura \ref{fig:frequency}) é um jogo acessível pensado para pessoas com deficiência visual e desenvolvido por \cite{ULF}. Esse jogo foi incluído por ter sido desenvolvido tendo a acessibilidade como foco principal. Todos os jogos foram baixados na sua versão iOS.

Os autores deste trabalho foram responsáveis pela avaliação. Ambos jogaram individualmente todos os jogos avaliados para compreender as mecânicas de cada um e explorar diversos fluxos dos aplicativos, incluindo menus, configurações e diferentes níveis do jogo. Antes de iniciar a avaliação, realizaram uma leitura de todas as diretrizes que seriam utilizadas para um melhor entendimento das mesmas. A etapa de avaliação de cada uma das \textit{guidelines} foi conduzida em conjunto pelos avaliadores, devido à necessidade de discussão sobre o escopo onde seriam aplicadas, a forma como estavam sendo atendidas ou não, e se a interpretação estava alinhada.

A avaliação dos jogos pelas diretrizes das Tabelas \ref{tab:guidelines_gag}, \ref{tab:guidelines_includification} e \ref{tab:guidelines_cognitive} foi feita considerando se o jogo atende a diretriz (sim), não atende (não), ou não se aplica (NA). A necessidade de uma opção ``não se aplica'' foi definida ao perceber que algumas \textit{guidelines} não estão presentes em alguns jogos mas não são relevantes para o contexto. Um exemplo desse cenário é o caso do jogo Tetris que não possui legendas, mas também não possui falas ou narração que necessite ser legendada. Ao realizar a avaliação pelas diretrizes do \cite{GAG} (Tabela \ref{tab:test_gag}), percebeu-se que por se tratarem de \textit{guidelines} mais gerais em termos de plataforma, algumas não se enquadravam no contexto de dispositivos móveis ou precisavam ser adaptadas pelos avaliadores na hora de considerar se o jogo atendia ou não. Enquanto a diretriz ``evita gatilhos de enjoo em simulação de VR'' não se aplicava para os jogos avaliados, a diretriz ``inclui uma opção para ajustar a sensibilidade dos controles'' precisou ser analisada pelos avaliadores que a forma de controle dos jogos era o toque na tela. 

Além das deficiências motora, auditiva, visual, cognitiva e de fala, o \cite{GAG} também traz uma categoria geral. Nessa categoria existem diretrizes que abrangem mais do que um tipo de deficiência, como por exemplo ``oferece uma ampla escolha de níveis de dificuldade'' e ``certifica-se de que todas as configurações sejam salvas/lembradas''. Também estão presentes nessa categoria diretrizes que falam a respeito das informações sobre acessibilidade, como incluir detalhes dos recursos de acessibilidade na embalagem ou no site e solicitar \textit{feedback} a respeito da acessibilidade do jogo. Dos três trabalhos utilizados para avaliação dos jogos, este foi o único que trouxe \textit{guidelines} informativas.

O jogo Frequency Missing foi desenvolvido pensando em atender as necessidades das pessoas com deficiência visual \cite{ULF}. Isso pode ser percebido ao longo da avaliação na medida em que todas as diretrizes trazidas na Tabela \ref{tab:test_gag} que se enquadram na categoria de deficiência visual e se aplicam ao jogo são atendidas. Todavia, nem todas as \textit{guidelines} das demais categorias são atendidas, mostrando a importância das decisões tomadas pela equipe de desenvolvimento referentes ao contexto do jogo: por ser um jogo especificamente pensado para deficiência visual, algumas diretrizes para deficiência motora, auditiva e cognitiva não foram implementadas.

\begin{small}
\begin{longtable}{|l|l|l|l|l|l|l|}
\caption{Avaliação pelas \textit{guidelines} de \cite{GAG}}
\label{tab:test_gag}\\
\hline
\textbf{\textit{Guideline}} & \textbf{Deficiência} & \textbf{\makecell[l]{Sub.\\Surf.}} & \textbf{Tetris} & \textbf{\makecell[l]{Freq.\\Missing}} & \textbf{\makecell[l]{Call\\of Duty}} & \textbf{\makecell[l]{Asph. 9}} \\ \hline
\endhead
\makecell[l]{Permite que os controles \\ sejam remapeados /\\reconfigurados.} & Motora & não & não & não & sim & sim \\ \hline
\makecell[l]{Certifica-se de que todas \\ as áreas da interface do \\ usuário possam ser acessadas \\ usando o mesmo método de \\ entrada do jogo.} & Motora & sim & sim & sim & sim & sim \\ \hline
\makecell[l]{Inclui uma opção para \\ ajustar a sensibilidade \\ dos controles.} & Motora & não & sim & não & sim & sim \\ \hline
\makecell[l]{Garante que os controles \\ sejam tão simples quanto \\ possível ou fornece uma \\ alternativa mais simples.} & Motora & sim & sim & sim & sim & sim \\ \hline
\makecell[l]{Garante que os elementos \\ interativos / controles virtuais\\  sejam grandes e bem \\ espaçados, especialmente em \\ telas pequenas ou sensíveis \\ ao toque.} & Motora & não & sim & sim & não & não \\ \hline
\makecell[l]{Inclui \textit{toggle} para \\ qualquer sensação tátil.} & Motora & não & sim & não & sim & sim  \\ \hline
\makecell[l]{Permite que o jogo seja \\ iniciado sem a necessidade \\ de navegar por vários níveis \\ de menus.} & Cognitiva & sim & sim & sim & não & não \\ \hline
\makecell[l]{Usa um tamanho de fonte \\ padrão facilmente legível.} & Cognitiva & não & sim & sim & não & não \\ \hline
\makecell[l]{Usa uma linguagem simples \\ e clara.} & Cognitiva & não & não & sim & não & sim \\ \hline
\makecell[l]{Usa formatação simples \\ de texto.} & Cognitiva & não & sim & sim & não & sim \\ \hline
\makecell[l]{Inclui tutoriais interativos.}  & Cognitiva & sim & não & sim & sim & sim \\ \hline
\makecell[l]{Permite que os jogadores \\ progridam através de avisos \\ de texto em seu próprio \\ ritmo.} & Cognitiva & sim & sim & não & não & sim \\ \hline
\makecell[l]{Evita imagens piscantes e \\ padrões repetitivos.} & Cognitiva & não & não & sim & não & não \\ \hline
\makecell[l]{Certifica-se de que nenhuma \\ informação essencial seja \\ transmitida apenas \\ por cor.}  & Visual & não & sim & sim & sim & sim \\ \hline
\makecell[l]{Se o jogo usar campo de \\ visão (somente mecanismo \\ 3D), define um padrão \\ apropriado para o ambiente \\ de visualização esperado.}  & Visual & NA & NA & NA & sim & sim \\ \hline
\makecell[l]{Evita gatilhos de enjoo \\ em simulação de VR.} & Visual & NA & NA & NA & NA & NA \\ \hline
\makecell[l]{Usa um tamanho de fonte \\ padrão facilmente legível.} & Visual & não & sim & sim & não & sim \\ \hline
\makecell[l]{Usa formatação simples \\ de texto.} & Visual & não & sim & sim & não & sim \\ \hline
\makecell[l]{Fornece alto contraste entre \\ texto / UI e plano de fundo.} & Visual & não & sim & sim & sim & sim \\ \hline
\makecell[l]{Garante que os elementos \\ interativos / controles \\ virtuais sejam grandes \\ e bem espaçados, \\especialmente em telas \\ pequenas ou sensíveis \\ ao toque.} & Visual & não & sim & sim & não & sim \\ \hline
\makecell[l]{Fornece legendas para todos \\ os discursos importantes.} & Auditiva & NA & NA & sim & NA & NA \\ \hline
\makecell[l]{Fornece controles de volume \\ separados ou silenciamentos\\  para efeitos, fala e \\ fundo / música.} & Auditiva & sim & sim & não & sim & sim \\ \hline
\makecell[l]{Certifica-se de que nenhuma \\ informação essencial seja \\ transmitida apenas por sons.} & Auditiva & sim & sim & sim & sim & sim \\ \hline
\makecell[l]{Se forem usadas legendas, \\ apresente-as de forma \\  clara e fácil de ler.} & Auditiva & NA & NA & sim & NA & NA \\ \hline
\makecell[l]{Certifica-se de que a entrada \\ de voz não seja necessária e \\ incluída apenas como um \\ método de entrada \\ complementar / alternativo.} & Fala & sim & sim & sim & sim & sim \\ \hline
\makecell[l]{Oferece uma ampla escolha \\ de níveis de dificuldade.} & Geral & não & não & não & sim & sim \\ \hline
\makecell[l]{Fornece detalhes dos \\ recursos de acessibilidade \\ na embalagem e / ou site.} & Geral & não & não & não & não & não \\ \hline
\makecell[l]{Fornece detalhes dos \\ recursos de acessibilidade \\ no jogo.} & Geral & não & não & sim & não & não\\ \hline
\makecell[l]{Certifica-se de que todas as \\ configurações sejam \\ salvas / lembradas.} & Geral & sim & sim & NA & sim & sim \\ \hline
\makecell[l]{Solicita \textit{feedback} \\ sobre acessibilidade.} & Geral & não & não & não & não & não \\ \hline
\end{longtable}
\end{small}

Na avaliação feita através das diretrizes propostas pelo \textit{Includification} \cite{ABLEGAMERS} (Tabela \ref{tab:test_includification}), foram consideradas apenas aquelas referentes ao contexto \textit{mobile}, embora o guia forneça mais diretrizes gerais. A escolha por considerar somente esse grupo na avaliação se deu pelo fato de que as demais \textit{guidelines} se assemelhavam às \textit{guidelines} já vistas no \textit{Game Accessibility Guidelines} \cite{GAG}, não havendo necessidade para repetição. As diretrizes trazidas na Tabela \ref{tab:guidelines_includification} estão descritas resumidamente, pois suas versões completas no guia são mais detalhadas, explicando até mesmo o motivo por trás de sua necessidade.

A análise dos jogos por essas diretrizes foi mais fácil de ser feita, uma vez que não precisou ser discutido se a diretriz era relevante para o contexto de dispositivos móveis ou não, pois todas já consideravam esse contexto. Por já estarem adaptadas para \textit{mobile}, não ficou margem para tantas dúvidas com relação a como interpretar a diretriz. Foi o caso da diretriz sobre botões alternativos, que deixa claro quais são os recursos especiais mencionados dando exemplos práticos em dispositivos móveis. Todavia, mesmo que todas as \textit{guidelines} dessa avaliação sejam pertinentes para jogos \textit{mobile}, algumas ainda não se aplicavam a todos os jogos avaliados, como é o caso da diretriz ``se o seu jogo exigir pressionar vários lugares ao mesmo tempo, considere agrupá-los'', onde apenas os jogos Call of Duty: Mobile e Asphalt 9 exigiam multi-toque.

\begin{small}
\begin{longtable} {|l|l|l|l|l|l|l|}
\caption{Avaliação pelas \textit{guidelines mobile} de \cite{ABLEGAMERS}}
\label{tab:test_includification} \\
\hline
\textbf{Categoria} & \textit{\textbf{Guideline}} & \textbf{\makecell[l]{Sub.\\Surf.}} & \textbf{Tetris} & \textbf{\makecell[l]{Freq.\\Missing}} & \textbf{\makecell[l]{Call\\of Duty}} & \textbf{\makecell[l]{Asph. 9}} \\ \hline 
\endhead
Toque & \makecell[l]{Grande área que pode \\ ser considerada a área \\ para tocar na tela.} & sim & sim & sim & não & não \\ \hline
Multi-toque & \makecell[l]{Se o seu jogo exigir \\ pressionar vários lugares \\ ao mesmo tempo, \\ considere agrupá-los.} & NA & NA & NA & não & sim \\ \hline
\makecell[l]{Botões\\alternativos} & \makecell[l]{Se um jogo exigir um \\ recurso especial, como \\ tocar na parte traseira \\ do dispositivo ou girá-lo, \\ permita que meios \\ alternativos sejam usados \\ para atingir o mesmo \\ objetivo.} & NA & NA & NA & NA & sim \\ \hline
\makecell[l]{Alto\\contraste} & \makecell[l]{Use cores que sejam \\ fáceis de distinguir \\ de outros elementos \\ ambientais do jogo.} & sim & sim & sim & sim & sim \\ \hline
& \makecell[l]{Opções de modo \\ daltonismo que podem ser\\ habilitadas.} & não & não & não & sim & não \\ \cline{2-7} 
\multirow{-2}{*}{\makecell[l]{Opções modo\\daltonismo}}  & \makecell[l]{Incluir símbolos para \\ distinguir elementos \\ visualmente.} & NA & sim & NA & sim & sim \\ \hline
\makecell[l]{Configuração\\de velocidade} & \makecell[l]{Capacidade de \\ desacelerar o jogo.} & não & não & não & não & não \\ \hline
\end{longtable}
\end{small}

As diretrizes de \cite{COGNITIVE} são propostas para jogos \textit{mobile} com foco apenas em deficiência cognitiva. Algumas \textit{guidelines} apresentadas já haviam sido vistas em \cite{GAG} ou em \cite{ABLEGAMERS}, uma vez que o trabalho utiliza ambos como referência. De maneira geral, não foram percebidas dificuldades ao avaliar os jogos escolhidos por essas diretrizes, e os resultados podem ser vistos na Tabela \ref{tab:test_cognitive}. Contudo, percebeu-se que algumas diretrizes estavam muito diretas, como a diretriz ``legendas'', dando a impressão de que legendas são obrigatórias em todos os cenários e não somente quando são relevantes. 

O modelo de avaliação de \cite{COGNITIVE} permite ordenamento quantitativo de jogos avaliados através da aplicação de um somatório que leva em consideração se dado jogo cumpre ou não com determinada diretriz. Tal processo foi adaptado durante a avaliação, onde foi decidido pela utilização de uma avaliação qualitativa a fim de evitar penalização por não cumprimento de alguma diretriz cuja relevância para o jogo em específico não existe. Sendo assim, embora o modelo proposto pelos autores avalie com uma soma, para fins de padronização e comparação neste trabalho, seguem os mesmos parâmetros definidos anteriormente de ``sim'', ``não'' e ``NA''.

\begin{small}
\begin{longtable} {|l|l|l|l|l|l|l|}
\caption{Avaliação pelas \textit{guidelines} de \cite{COGNITIVE}}
\label{tab:test_cognitive} \\
\hline
\textbf{Nível} & \textit{\textbf{Guideline}} & \textbf{\makecell[l]{Sub.\\Surf.}} & \textbf{Tetris} & \textbf{\makecell[l]{Freq.\\Missing}} & \textbf{\makecell[l]{Call\\of Duty}} & \textbf{\makecell[l]{Asph. 9}}\\ \hline
\endhead
\multirow{2}{*}{Baixo} & Utiliza uma linguagem simples. & não & sim & sim & não & sim   \\ \cline{2-7}
& \makecell[l]{Fontes personalizáveis\\(cor, tamanhos).}        &  não & não & não & não & não                                     \\ \cline{2-7}
& \makecell[l]{Ligar / desligar elementos\\gráficos.}           &  não & não & não & não & sim                                      \\ \cline{2-7}
& Legendas.                                      &  NA & NA & sim & NA & NA                                      \\ \cline{2-7}
& Progressão simples a difícil.                  &  não & não & não & não & sim                                      \\ \cline{2-7}
& Menus acessíveis.                              &  não & não & não & não & não                                      \\ \cline{2-7}
& Modos \textit{sandbox}.                                 &  não & não & não & sim & não                                      \\ \hline
\multirow{8}{*}{Médio}  & Níveis de treinamento. & não & não & não & sim & sim \\ \cline{2-7}
& \makecell[l]{Lembrança dos objetivos durante\\ o jogo.}          &  não & sim & sim & sim & sim                                      \\ \cline{2-7}
& \makecell[l]{Lembrança dos controles durante\\ o jogo.}          &  não & não & não & sim & sim                                      \\ \cline{2-7}
& \makecell[l]{Configuração alternativa de\\arquivos de som.}   &  sim & não & não & sim & não                                      \\ \cline{2-7}
& \makecell[l]{Use recompensas visuais\\explícitas.}            &  sim & sim & não & sim & sim                                      \\ \cline{2-7}
& Possibilidade de repetição.                    &  sim & não & sim & sim & sim                                      \\ \cline{2-7}
& \makecell[l]{Pausa enquanto o texto está\\sendo lido.}        &  não & sim & não & não & sim                                      \\ \cline{2-7}
& Salvar configurações.                          &  sim & sim & NA & sim & Sim                                      \\ \hline
\multirow{5}{*}{Alto} & \makecell[l]{Mira automática, capacidade \\ de travar um alvo.} & NA & NA & NA & sim & sim \\ \cline{2-7}
& Configurações de velocidade.                   &  não & não & NA & não & não                                      \\ \cline{2-7}
& Capacidade de \textit{voice over}.             &  não & não & sim & não & não                                      \\ \cline{2-7}
& \makecell[l]{Sensibilidade ajustável / \\tolerância a erros.}  &  não & sim & não & sim & sim                                      \\ \cline{2-7}
& Passe automático.                              &  NA & não & não & não & não                                      \\ \hline
\end{longtable}
\end{small}

\subsection{Proposta de um novo conjunto de \textit{guidelines}} \label{proposta}

A partir da análise das diretrizes utilizadas para avaliação, este trabalho propõe um novo conjunto de \textit{guidelines} para desenvolvimento de jogos \textit{mobile}, que podem ser encontradas na Tabela \ref{tab:guidelines}. Para definição do escopo abrangido, algumas diretrizes dos trabalhos relacionados não foram incluídas por possuírem relevância em contextos mais específicos. Um exemplo é a diretriz que fala sobre evitar gatilhos de enjoo em simulação de realidade virtual, presente em \cite{GAG}, cuja aplicabilidade foge do escopo deste trabalho devido ao \textit{hardware} adicional necessário e foi, portanto, desconsiderada durante a criação das \textit{guidelines} propostas. 

Foram consideradas também as práticas sugeridas pela Apple\footnote{https://developer.apple.com} na sua documentação de acessibilidade \cite{HIG}. Essas \textit{guidelines} são diretamente estruturadas para serem utilizadas no desenvolvimento de aplicativos para dispositivos móveis. 
Embora não sejam especificamente para jogos, elas tratam de questões pertinentes ao contexto \textit{mobile} como a capacidade de leitor de tela e o ajuste do tamanho da fonte. Esses dois comportamentos já estão presentes de forma nativa dentro do sistema operacional dos dispositivos, podendo apenas serem integradas e configuradas dentro do jogo no momento do desenvolvimento.

Com relação ao nível das diretrizes, a maior parte está inserida dentro do nível básico. Esse nível foi definido levando em consideração a facilidade de implementação e adaptação do jogo para que a diretriz seja atendida. As \textit{guidelines} propostas servem como orientações para serem seguidas levando em consideração cada caso em específico. Assim, para cenários onde uma diretriz não se aplica, não há necessidade de implementação. 

\begin{small}
\begin{longtable}{ |l|l|l|l| }
\caption{\textit{Guidelines} propostas}
\label{tab:guidelines} \\
\hline
\textbf{\textit{Guideline}} & \textbf{Deficiência} & \textbf{Referência} & \textbf{Estado} \\
\hline
\endhead

\makecell[l]{Nenhuma informação importante é \\ representada apenas por sons.} & Auditiva & \cite{GAG} & Reescrita \\ \hline
\makecell[l]{Quando necessário, apresentar \\ legendas.} & Auditiva & \cite{ABLEGAMERS} & Reescrita \\ \hline
\makecell[l]{Efeitos sonoros também são \\ representados visualmente.} & Auditiva & \cite{ABLEGAMERS} & Reescrita \\ \hline
\makecell[l]{Sem limite de tempo para ler textos.} & Cognitiva & \makecell[l]{[Jaramillo-Alcázar\\et al. 2017]} & Reescrita \\ \hline
\makecell[l]{Permite que o jogo seja iniciado \\ sem a necessidade de navegar por \\ vários níveis de menus.} & Cognitiva & \cite{GAG} & Original \\ \hline
\makecell[l]{Opção de mira automática ou fixar \\ no alvo.} & Cognitiva & \makecell[l]{[Jaramillo-Alcázar\\et al. 2017]} & Original \\ \hline
\makecell[l]{Utiliza uma linguagem simples e clara.} & Cognitiva & \cite{GAG} & Original \\ \hline
\makecell[l]{Não utiliza muitos efeitos especiais \\ visuais repetitivos.} & Cognitiva & \cite{GAG} & Reescrita \\ \hline
\makecell[l]{Tutoriais interativos de como jogar.} & Cognitiva & \makecell[l]{[Jaramillo-Alcázar\\et al. 2017] \\ \cite{GAG}} & Reescrita \\ \hline
\makecell[l]{Lembrança dos objetivos durante \\ o jogo.} & Cognitiva &\makecell[l]{[Jaramillo-Alcázar\\et al. 2017]} & Original \\ \hline
\makecell[l]{Lembrança dos comandos durante \\ o jogo.} & Cognitiva & \makecell[l]{[Jaramillo-Alcázar\\et al. 2017]} & Original \\ \hline
\makecell[l]{Apresenta possibilidade de \\ repetição.} & Cognitiva & \makecell[l]{[Jaramillo-Alcázar\\et al. 2017]} & Reescrita \\ \hline
\makecell[l]{Não requer precisão do toque.} & Motora & \cite{ABLEGAMERS} & Adaptada \\ \hline
\makecell[l]{Reação rápida não é obrigatória.} & Motora & \cite{ABLEGAMERS} & Original \\ \hline
\makecell[l]{Tempo para realizar os comandos \\ não é importante.} & Motora & \cite{ABLEGAMERS} & Adaptada \\ \hline
\makecell[l]{Permitir jogar no modo paisagem \\ ou retrato.} & Motora & \cite{GAG} & Original \\ \hline
\makecell[l]{Elementos clicáveis são bem \\ espaçados na tela e possuem um \\ bom tamanho.} & Motora & \cite{GAG} & Adaptada \\ \hline
\makecell[l]{Caso sejam usados comandos \\ especiais (toque atrás do dispositivo, \\ rotacionar, etc.), fornecer uma \\ alternativa.} & Motora & \cite{ABLEGAMERS} & Reescrita \\ \hline
\makecell[l]{Se necessário utilizar multi-toque, \\ fornecer uma alternativa.} & Motora & \cite{ABLEGAMERS} & Reescrita \\ \hline
\makecell[l]{Adaptação ao tamanho da fonte \\ selecionada no dispositivo (Dynamic \\ Type na plataforma iOS).} & Visual & \cite{HIG} & Reescrita \\ \hline
\makecell[l]{Capacidade de leitor de tela \\ (VoiceOver na plataforma iOS, \\ TalkBack na plataforma Android, \\ etc.).} & Visual & \cite{HIG} & Reescrita \\ \hline
\makecell[l]{Opções de cores de modo daltonismo.} & Visual & \cite{ABLEGAMERS} & Reescrita \\ \hline
\makecell[l]{Nenhuma informação importante \\ é representada apenas por cores.} & Visual & \cite{ABLEGAMERS} & Reescrita \\ \hline
\makecell[l]{Alto contraste entre elementos \\ visuais e o fundo.} & Visual & \cite{ABLEGAMERS} & Reescrita \\ \hline
\makecell[l]{Comandos de voz não são \\ obrigatórios.} & Fala & \cite{GAG} & Reescrita \\ \hline
\makecell[l]{Usa formatação simples de texto.} & \makecell[l]{Cognitiva \\ e Visual} & \cite{GAG} & Original \\ \hline
\makecell[l]{Possibilidade de ajustar o volume \\ dos efeitos sonoros, narração e sons \\ ambientes separadamente.} & \makecell[l]{Auditiva \\ e Cognitiva} & \cite{GAG} & Reescrita \\ \hline
\makecell[l]{Sensibilidade do toque pode ser \\ ajustada.} & \makecell[l]{Motora \\ e Cognitiva} & \makecell[l]{$[$Bartlet and Spohn 2012$]$ \\ $[$Jaramillo-Alcázar \\ et al. 2017$]$} & Adaptada \\ \hline
\makecell[l]{Possibilidade de habilitar \textit{feedbacks} \\ hápticos.} & \makecell[l]{Motora \\e Visual} & \makecell[l]{\cite{HIG} \\ \cite{GAG}} & Reescrita \\ \hline
\makecell[l]{Opções de diferentes níveis de \\ dificuldade para serem selecionadas.} & \makecell[l]{Cognitiva \\e Geral} & \cite{GAG} & Reescrita \\ \hline
\makecell[l]{Menus fáceis de acessar.} & \makecell[l]{Cognitiva \\ e Geral} & \cite{ABLEGAMERS} & Reescrita \\ \hline
\makecell[l]{Usa sons distintos para diferentes \\ objetos e eventos.} & \makecell[l]{Cognitiva \\ e Visual} & \cite{GAG} & Adaptada \\ \hline
\makecell[l]{Garantia de que todas as configurações \\ são salvas.} & Geral & \cite{ABLEGAMERS} & Original \\ \hline
\end{longtable}
\end{small}

As \textit{guidelines} estão categorizadas de acordo com o tipo de deficiência e seguem a mesma categoria na qual a diretriz do trabalho de referência estava colocada, abrangendo deficiência auditiva, cognitiva, motora, visual e de fala. Ademais, cada diretriz possui um estado: original, reescrita ou adaptada. O estado original se refere às diretrizes que foram apenas traduzidas do idioma no qual se encontravam para o Português. Já as reescritas são aquelas que sofreram alguma alteração no texto mas o seu significado permanece sem alterações. Essas alterações foram feitas para resumir alguma diretriz que estava muito longa ou para explicar de uma forma mais clara de acordo com a opinião dos autores. Por fim, as adaptadas sofreram modificações do seu estado original principalmente para melhor atender ao contexto dos jogos para dispositivos móveis. Três das quatro adaptações feitas se devem pela principal forma de comando nesses dispositivos se dar através do toque na tela, diferente de jogos para console que utilizam controles ou \textit{joystick}. Essas \textit{guidelines} estão destacadas na Tabela \ref{tab:guidelines_adaptadas}, junto como a sua versão original e o motivo da adaptação.

\begin{small}
\begin{longtable}{ |l|l|l| }
\caption{\textit{Guidelines} adaptadas}
\label{tab:guidelines_adaptadas} \\
\hline
\textbf{\textit{Guideline}} & \textbf{Original} & \textbf{Justificativa} \\
\hline
\endhead

\makecell[l]{Não requer precisão do \\ toque.} & \makecell[l]{Não requer precisão. \\ \cite{ABLEGAMERS}} & \makecell[l]{Adaptada para o contexto \\ \textit{mobile} onde é utilizado \\ o toque.} \\ \hline
\makecell[l]{Tempo para realizar os \\comandos não é importante.} & \makecell[l]{Tempo de movimento / apertar \\ botões não é importante. \\ \cite{ABLEGAMERS}} & \makecell[l]{Adaptada para o contexto \\ \textit{mobile} que raramente \\ usa botões.} \\ \hline
\makecell[l]{Elementos clicáveis são \\ bem espaçados na tela e \\ possuem um bom tamanho.} & \makecell[l]{Garante que elementos \\ interativos / controles virtuais \\ são grandes e bem espaçados, \\ especialmente em telas \\ pequenas ou de toque. \\ \cite{GAG}} & \makecell[l]{Sem necessidade de \\ especificar o tipo de tela \\ no contexto \textit{mobile}.} \\ \hline
\makecell[l]{Sensibilidade do toque \\ pode ser ajustada.} & \makecell[l]{Sensibilidade da câmera / \textit{joystick}. \\ $[$Bartlet and Spohn 2012$]$ \\ Sensibilidade / tolerância de \\ erro ajustável. \\ $[$Jaramillo-Alcázar et al. 2017$]$} & \makecell[l]{Adaptada para o contexto \\ \textit{mobile} onde é utilizado \\ o toque.} \\ \hline
\end{longtable}
\end{small}

\section{Resultados} \label{resultados}

Após a definição do novo conjunto de diretrizes propostas, foi realizada uma nova avaliação dos jogos Subway Surfers, Tetris, Call of Duty: Mobile, Asphalt 9 e Frequency Missing com estas \textit{guidelines}. A mesma forma de avaliação foi utilizada, considerando se o jogo atende a diretriz (sim), não atende (não) ou não se aplica (NA). Embora todas as \textit{guidelines} sejam pertinentes para o contexto de jogos para dispositivos móveis, nem todas são relevantes para o tipo de jogo sendo avaliado, fazendo com que a opção ``NA'' continue sendo necessária.

\begin{small}
\begin{longtable}{ |l|l|l|l|l|l|l| }
\caption{Avaliação pelas \textit{guidelines} propostas}
\label{tab:test_guidelines} \\
\hline
\textit{\textbf{\makecell[l]{Guideline}}} & \textbf{\makecell[l]{Categoria}} & \textbf{\makecell[l]{Sub.\\Surf.}} & \textbf{Tetris} & \textbf{\makecell[l]{Freq.\\Missing}} & \textbf{\makecell[l]{Call\\of Duty}} & \textbf{Asph. 9} \\
\hline
\endhead

\makecell[l]{Nenhuma informação\\importante é representada\\apenas por sons.} & \makecell[l]{Auditiva} & sim & sim & sim & sim & sim \\ \hline
\makecell[l]{Quando necessário\\apresentar legendas.} & \makecell[l]{Auditiva} & NA & NA & sim & NA & NA \\ \hline
\makecell[l]{Efeitos sonoros também\\são representados\\visualmente.} & \makecell[l]{Auditiva} & sim & sim & sim & sim & sim \\ \hline
\makecell[l]{Sem limite de tempo para\\ler textos.} & \makecell[l]{Cognitiva} & sim & sim & não & não & sim \\ \hline
\makecell[l]{Permite que o jogo seja\\iniciado sem a necessidade\\de navegar por vários\\níveis de menus.} & \makecell[l]{Cognitiva} & sim & sim & sim & não & não \\ \hline
\makecell[l]{Opção de mira automática\\ou fixar no alvo.} & \makecell[l]{Cognitiva} & NA & NA & NA & sim & sim \\ \hline
\makecell[l]{Utiliza uma linguagem\\simples e clara.} & \makecell[l]{Cognitiva} & não & não & sim & não & sim \\ \hline
\makecell[l]{Não utiliza muitos efeitos\\especiais visuais repetitivos.} & \makecell[l]{Cognitiva} & não & não & sim & não & não \\ \hline
\makecell[l]{Tutoriais interativos de\\como jogar.} & \makecell[l]{Cognitiva} & sim & não & sim & sim & sim \\ \hline
\makecell[l]{Lembrança dos objetivos\\durante o jogo.} & \makecell[l]{Cognitiva} & não & sim & sim & sim & sim \\ \hline
\makecell[l]{Lembrança dos comandos\\durante o jogo.} & \makecell[l]{Cognitiva} & não & não & não & sim & sim \\ \hline
\makecell[l]{Apresenta possibilidade\\de repetição.} & \makecell[l]{Cognitiva} & sim & não & sim & sim & sim \\ \hline
\makecell[l]{Não requer precisão\\do toque.} & \makecell[l]{Motora} & sim & sim & não & não & sim \\ \hline
\makecell[l]{Reação rápida não é\\obrigatória.} & \makecell[l]{Motora} & não & não & sim & não & não \\ \hline
\makecell[l]{Tempo para realizar os\\comandos não é importante.} & \makecell[l]{Motora} & não & não & sim & não & não \\ \hline
\makecell[l]{Permitir jogar no modo\\paisagem ou retrato.} & \makecell[l]{Motora} & não & não & não & não & não \\ \hline
\makecell[l]{Elementos clicáveis são\\bem espaçados na tela e\\possuem um bom tamanho.} & \makecell[l]{Motora} & não & sim & sim & não & não \\ \hline
\makecell[l]{Caso sejam usados comandos\\especiais (toque atrás do\\dispositivo, rotacionar, etc.), \\fornecer uma alternativa.} & \makecell[l]{Motora} & NA & NA & NA & NA & sim \\ \hline
\makecell[l]{Se necessário utilizar\\multi-toque, fornecer uma\\alternativa.} & \makecell[l]{Motora} & NA & NA & NA & não & sim \\ \hline
\makecell[l]{Adaptação ao tamanho da\\fonte selecionada no\\dispositivo (Dynamic Type\\na plataforma iOS).} & \makecell[l]{Visual} & não & não & não & não & não \\ \hline
\makecell[l]{Capacidade de leitor\\de tela (VoiceOver na\\plataforma iOS, TalkBack\\na plataforma Android, etc.).} & \makecell[l]{Visual} & não & não & sim & não & não \\ \hline
\makecell[l]{Opções de cores de modo\\daltonismo.} & \makecell[l]{Visual} & não & não & não & sim & não \\ \hline
\makecell[l]{Nenhuma informação\\importante é representada\\apenas por cores.} & \makecell[l]{Visual} & sim & sim & sim & sim & sim \\ \hline
\makecell[l]{Alto contraste entre\\elementos visuais e o\\fundo.} & \makecell[l]{Visual} & não & sim & sim & sim & sim \\ \hline
\makecell[l]{Comandos de voz não são\\obrigatórios.} & \makecell[l]{Fala} & sim & sim & sim & sim & sim \\ \hline
\makecell[l]{Usa formatação simples\\de texto.} & \makecell[l]{Cognitiva\\e Visual} & não & não & sim & não & sim \\ \hline
\makecell[l]{Possibilidade de ajustar\\o volume dos efeitos sonoros,\\narração e sons ambientes\\separadamente.} & \makecell[l]{Auditiva\\e Cognitiva} & sim & sim & NA & sim & sim \\ \hline
\makecell[l]{Sensibilidade do toque\\pode ser ajustada.} & \makecell[l]{Motora\\e Cognitiva} & não & sim & não & sim & sim \\ \hline
\makecell[l]{Possibilidade de habilitar\\\textit{feedbacks} hápticos.} & \makecell[l]{Motora\\e Visual} & não & sim & não & sim & sim \\ \hline
\makecell[l]{Opções de diferentes níveis\\de dificuldade para serem \\selecionadas.} & \makecell[l]{Cognitiva\\e Geral} & NA & não & não & sim & sim \\ \hline
\makecell[l]{Menus fáceis de acessar.} & \makecell[l]{Cognitiva\\e Geral} & não & não & sim & não & não \\ \hline
\makecell[l]{Usa sons distintos para\\diferentes objetos e eventos.} & \makecell[l]{Cognitiva\\e Visual} & sim & sim & sim & sim & sim \\ \hline
\makecell[l]{Garantia de que todas as\\configurações são salvas.} & \makecell[l]{Geral} & sim & sim & sim & sim & sim \\ \hline
\end{longtable}
\end{small}

A necessidade de manter uma opção de ``NA'' se evidencia na diretriz ``quando necessário apresentar legendas'': todos os jogos avaliados anteriormente como não se aplica mantiveram o mesmo resultado. Isso acontece pois a interpretação da \textit{guideline} proposta conservou o significado da \textit{guideline} original, onde jogos sem diálogos e/ou narração continuam não possuindo conteúdo a ser legendado. Esse processo de interpretação livre das diretrizes teve sua importância notada durante a avaliação feita na metodologia e foi levado em consideração na criação da Tabela \ref{tab:guidelines}, onde diretrizes buscam referenciar em linhas gerais um caminho a ser seguido ao invés de estipular explicitamente o que deve ser feito. 

O caráter orientador das \textit{guidelines} propostas na Tabela \ref{tab:guidelines} pôde ser observado durante a avaliação do Subway Surfers. Anteriormente o jogo não havia cumprido com a  \textit{guideline} ``pausa para ler textos'' de \cite{COGNITIVE}, devido à necessidade do jogo ser pausado. No entanto, a  \textit{guideline} proposta ``sem limite de tempo para ler textos'' traz liberdade no processo de \textit{game design} sem abrir mão da inclusão. 
Durante o tutorial, o jogo utiliza artifícios de desaceleração e repetição dos elementos a fim de garantir a presença dos textos informativos na tela até interação do usuário para avançar até o próximo passo, cumprindo com a diretriz proposta da sua própria maneira.

As \textit{guidelines} ``não requer precisão do toque'' e ``sensibilidade do toque pode ser ajustada'' foram propostas trazendo menos ambiguidade de resultados quando comparada à original ``sensibilidade ajustável / tolerância a erros'' presente em \cite{COGNITIVE}. A separação em duas diretrizes distintas permitiu uma obtenção de resultados mais diretos, clarificando qual âmbito foi ou não alcançado pelo jogo, como é no caso do Subway Surfers. Mesmo que possua tolerância a erros, o jogo foi anteriormente avaliado com ``não atende'' por conta da ausência de um ajuste de sensibilidade dos controles. A avaliação negativa por cumprimento parcial de uma \textit{guideline} que possui dois itens foi evitada com a separação dos mesmos em duas diretrizes distintas.

Por outro lado, Call of Duty: Mobile foi avaliado como ``sim'' anteriormente, pois a sensibilidade dos controles podem ser ajustadas e a tolerância a erros está presente: ainda que o alvo não esteja na mira, o projetil o atinge em decorrência do assistente de mira. Contudo, após separação e reescrita das \textit{guidelines}, acabou por receber resultado negativo no que tange a precisão do toque. O motivo foi a adaptação da \textit{guideline} original para o contexto \textit{mobile}, onde a precisão necessária no toque em uma tela se torna mais notável quanto menor o tamanho da tela do dispositivo e maior o número de alvos inimigos a serem eliminados durante o jogo, ainda que com o uso de assistente de mira.

Estas diferenças percebidas durante o processo de avaliação pelo novo conjunto de diretrizes propostas se dá pela maior especificação do contexto na qual elas estão aplicadas. Observa-se uma avaliação mais direcionada para jogos \textit{mobile}, eliminando a necessidade de adaptação. Além disso, a separação das diretrizes, que anteriormente abrangiam mais de uma condição, proporcionou uma análise mais precisa dos jogos avaliados. Isso permite uma compreensão mais detalhada dos pontos que são atingidos ou não, contribuindo para uma avaliação mais refinada e precisa.

\section{Considerações Finais} \label{consideracoes}

Sabendo da importância da cultura de jogos para as pessoas se sentirem inseridas na sociedade, é muito importante fazer com que cada vez mais os jogos sejam inclusivos. O uso de diretrizes como guia na hora de desenvolver jogos inclusivos serve como um ponto de começo pra estruturar o \textit{game design} levando em conta critérios de acessibilidade. Muitas vezes pequenos ajustes, como o cuidado com o contraste de cores e o uso de legendas quando necessário, podem fazer com que mais pessoas sejam incluídas e tenham suas necessidades atendidas. 

Dado seu caráter qualitativo, as diretrizes propostas neste trabalho precisam ser avaliadas de acordo com a necessidade e contexto de cada jogo desenvolvido. Elas servem tanto como auxílio na hora de estruturar o desenvolvimento de um novo jogo, como também uma forma de avaliação para jogos já existentes. Maiores especificações de cada diretriz, como forma de implementação ou escopo a ser atendido, precisam ser definidas pelo time responsável na hora de serem aplicadas.

Neste trabalho, as \textit{guidelines} propostas foram empregadas como instrumento para avaliar jogos já existentes no mercado. Para trabalhos futuros, recomenda-se a concepção de um jogo com base nessas diretrizes, desde a fase inicial de desenvolvimento até os testes com usuários. Essa abordagem possibilitará uma análise mais abrangente das diretrizes em relação ao ciclo completo de desenvolvimento, permitindo uma compreensão mais profunda de sua aplicabilidade e impacto ao longo do processo.

\bibliographystyle{sbc}
\bibliography{sbc-template}

\end{document}